\documentclass[a4paper,12pt]{article}
\pdfoutput=2
\usepackage{mathrsfs, subfigure}
\usepackage{amsmath}
\usepackage{amsfonts}
\usepackage{amssymb}
\usepackage{graphicx}
\usepackage{diagbox}
\usepackage{jheppub}

\newcommand{\be}{\begin{equation}}
\newcommand{\bea}{\begin{eqnarray}}
\newcommand{\eea}{\end{eqnarray}}
\newcommand{\ba}{\begin{array}}
\newcommand{\ea}{\end{array}}
\newcommand{\ee}{\end{equation}}

\def\bse{\begin{subequations}}
	\def\ese{\end{subequations}}

\title{Deforming charged black holes with dipolar differential rotation  boundary}
\author{Tong-Tong Hu,}
\author{Shuo Sun,}
\author{Hong-Bo Li and}
\author{Yong-Qiang Wang}
\affiliation{Research Center of Gravitation $\&$ Institute of Theoretical Physics, Lanzhou University, Lanzhou 730000, China}
\affiliation{Key Laboratory for Magnetism and Magnetic of the Ministry of Education, Lanzhou University, Lanzhou 730000, China}
\emailAdd{hutt17@lzu.edu.cn,sunsh17@lzu.edu.cn,lihb2017@lzu.edu.cn,
	yqwang@lzu.edu.cn}

\abstract{Motivated by the recent studies of the novel asymptotically  global AdS$_4$ black hole with deforming horizon, we consider the action of Einstein-Maxwell gravity in AdS spacetime and construct the charged deforming AdS black holes with differential boundary. In contrast to deforming black hole without charge, there exists at least one value of horizon for an arbitrary temperature. The extremum of temperature is determined by charge $q$ and divides the range of temperature into several parts. Moreover, we use an isometric embedding in the three-dimensional space  to investigate the horizon geometry.  We also study the entropy and quasinormal modes of deforming charged AdS black hole. It is interesting to find there exist two families of black hole solutions with different horizon radius for a fixed temperature, but these two black holes have same horizon geometry and entropy. Due to the existence of charge $q$, the phase diagram of entropy is more complicated. }

\keywords{AdS/CFT duality, Black holes, Quasinormal modes  }
\begin{document}
	\maketitle
	
	\section{Introduction}
	\hspace*{0.6cm} In classical general relativity, due to the uniqueness theorem of black holes \cite{a1,1,2,3}, the asymptotically flat charged  black hole solutions with zero angular momentum in four dimensions are named as Reissner-Nordstrom (RN)  black holes, which have two spherical event horizons. In four-dimensional anti-de Sitter (AdS) spacetime, it is well-known that except for compact horizons of arbitrary genus, there exist some solutions with noncompact planar or hyperbolic horizons. Because of the development of Anti-de Sitter/conformal field theory (AdS/CFT) correspondence  \cite{Maldacena:1997re,Maldacena:1998re,Witten:1998qj,Aharony:1999ti}, it becomes more important to study  physical properties of AdS black holes.
	\par
	Because the asymptotically AdS black hole has a boundary metric of conformal structure, we could deform the boundary metric to obtain a family of solutions of black hole with deforming horizon, whose curvature is not a constant value.
     There are many works in this field with both analytical and numerical methods. For the analytic method, the authors  in \cite{Chen:2015zoa} constructed a family of black hole solutions with deforming horizons in four-dimensional spacetime by using AdS C-metric \cite{levicivita1917,weyl1917,Plebanski:1976gy}. In addition, a class of solutions
	of four-dimensional AdS black holes with noncompact event horizons of finite area was
found in \cite{Klemm:2014rda,Gnecchi:2013mja},
and black holes with bottle-shaped event horizon were founded analytically in \cite{Chen:2016rjt}. With numerical methods, the authors in \cite{Markeviciute:2017jcp} got a family of deforming solutions including soliton and black hole with dipolar differential boundary $\Omega(\theta)=\varepsilon \cos(\theta)$. The constant  $\varepsilon$ is the boundary rotation parameter and $\theta$ is polar angle. When $\varepsilon>2$, the norm Killing vector $\partial _t$ becomes spacelike for certain regions which also are called as ergoregions, and deforming AdS black holes with ergoregions may be unstable due to superradiant scattering \cite{Green:2015kur}. Because of superradiance, both solitons and black holes develop hair at $\varepsilon>2$. Motivated by this work, we also studied deforming solutions with odd multipolar \cite{Sun:2019qsf} and even multipolar \cite{Li:2019tsm} differential
	rotation boundary. Furthermore, in \cite{Crisford:2018qkz}, the authors numerically studied a class of  vacuum solutions with a noncompact, differential rotation boundary metric. With AdS C-metric, the effect of changing boundary metric on hyperbolic and
	compact AdS black holes had been studied in \cite{Horowitz:2018coe}. Considering the matter fields, the authors in \cite{a17} constructed the deforming black holes in $D = 5$ minimal gauged supergravity.
	\par
	Until now, the works of deforming AdS black hole with differential boundary \cite{Markeviciute:2017jcp, Li:2019tsm,Sun:2019qsf} are only studied in the situations without charge. It would be interesting to see whether there exist the charged deforming AdS black holes solutions in Einstein-Maxwell-AdS spacetime. In this paper, we would like to numerically solve coupled Einstein-Maxwell equations to obtain a family of solutions of charged deforming black holes. These solutions have the anti-symmetric rotation profile on the equatorial plane, which keeps total angular momentum of black hole being zero. In contract to the situations without charge, there exist some new properties of black holes due to the existence of charge $q$. Firstly, there exists at least one value of horizon for an arbitrary temperature, while there exists no horizon when $T<T_{min}$ for $q=0$. Besides, the extremum of temperature is determined by charge $q$ and divide temperature into several parts according to the charge $q$. In different regions of temperature, the number of values for horizon is different. Specifically, in the region with one value of horizon for a fixed temperature, there exist two families of solutions with same horizon when temperature is lower than the minimal extremum of temperature for RN-AdS black hole $T_{RN}=\frac{\sqrt{6}}{3\pi}$. Furthermore, in the region with three values of horizon for a fixed temperature, it is interesting to find that two small branches have same properties such as horizon geometry, entropy and quasinormal modes, although their horizon radii are not equal.
	\par
	The plan of our work is as follows. In the next section, we introduce our model and numerical method. In Sec. \ref{Sec3}, we obtain numerical solutions of charged AdS black hole with differential rotation boundary and discuss the effect of the  temperature $T$ and the charge $q$ on solutions. We also show more properties of deforming charged AdS black hole including horizon geometry, entropy and stability. The conclusions and outlook are given in the last section.
\section{Model and numerical method}\label{Sec2}

 \hspace*{0.6cm} We start with Einstein-Maxwell action in four-dimensional AdS spacetime, whose action is given by
 \begin{align}
 S=\frac{1}{16\pi G}\int \mathrm{d}^4x&\sqrt{-g}\left(R-2\Lambda-\frac{1}{4}F^{\mu\nu}F_{\mu\nu}\right),
 \label{eq:action}
\end{align}
 where $G$ is the gravitational constant, $\Lambda$ is cosmological constant represented by AdS radius $L$ as $-3/L^2$ in four-dimensional spacetime, $g$ is determinant of metric and $R$ is Ricci scalar.
\par
The equations of motion of the Einstein and the Maxwell fields which can
be derived from the Lagrangian density (\ref{eq:action}) are as follows
\begin{subequations}\label{m}
\begin{equation}\label{equation1}
R_{\mu\nu}+\frac {3}{L^2}g_{\mu\nu}-(\frac {1}{2}F_{\mu\lambda}{F_{\nu}}^{\lambda}-\frac {1}{8}g_{\mu\nu}F_{\lambda\rho}F^{\lambda\rho})=0,
\end{equation}
\begin{equation}\label{equation2}
\nabla_{\mu}F^{\mu\nu}=0.
\end{equation}
\end{subequations}
The spherically symmetric solution of motion equations  (\ref{m}) is
the well-known Reissner-Nordstrom-AdS (RN-AdS) black hole. The metric of RN-AdS black hole solution  could be written as follows
\begin{eqnarray}\label{matric1}
 ds^{2} &=& -\left(1-\frac{2M}{r}+\frac{q^2}{r^2}+\frac{r^2}{L^2}\right)dt^2+\left(1-\frac{2M}{r}+\frac{q^2}{r^2}+\frac{r^2}{L^2}\right)^{-1}dr^2 +r^2 d\Omega^2,
 \label{eq:RN}
\end{eqnarray}
and the gauge field is written as
\begin{equation}\label{gauge}
F=dA, \;\;\; A=\frac{q}{r}dt.
\end{equation}
Here, $d\Omega^2$ represents the standard element on $S^2$,
 the constant  $M$ is the mass of black hole measured
from the infinite boundary, and  the constant $q$ is the charge of black hole.
The black hole
mass is related to the charge $q$ and horizon radius $r_+$ by the relation
\begin{equation}\label{root}
M=\frac{1}{2}\left(r_++\frac{q^2}{r_+}+\frac{r_+^3}{L^2} \right),
\end{equation}
where $r_{+}$ is the largest root. The Hawking
temperature $T_H$ of RN-AdS black hole is given by
\begin{eqnarray}
 T_H=\frac{1}{4\pi r_+} \left(1+ \frac{3 r_+^2}{L^2}-\frac{q^2}{r_+^2}\right).
\end{eqnarray}
At near infinity the metric is asymptotic to the anti-de Sitter spacetime, and boundary
metric is given by
\begin{equation}\label{boundary}
  ds_\partial^2=-dt^2+d\theta^2+\sin^2\theta d\phi^2.
\end{equation}
\par
 In order to obtain the new asymptotic Anti-de Sitter solution, the authors in \cite{Markeviciute:2017jcp} add differential rotation to the boundary
metric, which is given by
\begin{equation}
   ds_\partial^2=-dt^2+d\theta^2+\sin^2\theta[d\phi+\Omega(\theta)dt]^2,
\label{eq:boundary}
 \end{equation}
with  a dipolar differential rotation $\Omega(\theta)=\varepsilon\cos\theta$. The constant $\varepsilon$ is the amplitude of the boundary rotation. The norm of Killing vector $\partial_t$ is
\begin{equation}
   \|{\partial{t}}\|^2=-1+\frac{\varepsilon^2}{4}\sin^2({2\theta}).
   \label{eq:timelike}
 \end{equation}
From the above equation, we could find that the maximal value of Killing vector appears at $\theta=\frac{\pi}{4}$. We will take the same dipolar differential rotation boundary (\ref{eq:boundary}).
\par
 In order to get a set of charged deforming black hole solutions, we would like to use DeTurk method \cite{Headrick:2009pv,Wiseman2012,Dias:2015nua} to solve equations of motion (\ref{m}). By adding a gauge fixing term, we change equations (\ref{equation1}) to elliptic equations:
 \begin{equation}\label{DTequation}
 R_{\mu\nu}+\frac {3}{L^2}g_{\mu\nu}-(\frac {1}{2}F_{\mu\lambda}{F_{\nu}}^{\lambda}-\frac {1}{8}g_{\mu\nu}F_{\lambda\rho}F^{\lambda\rho})-\nabla_{(_\mu}\xi_{\nu)}=0,
\end{equation}
where the Deturk vector $\xi^\mu=g^{\nu\rho}(\Gamma^\mu_{\nu\rho}[g]-\Gamma^\mu_{\nu_\rho}[\tilde{g}])$ is related to reference metric $\tilde{g}$. It is notable that the reference metric $\widetilde{g}$ should be chosen to have the same boundary and horizon structure with $g$. Using this method to solve equations (\ref{equation2}) and (\ref{DTequation}), we could obtain a family of charged AdS black hole solutions.

\section{Black hole solutions}	\label{Sec3}

\hspace*{0.6cm}  To obtain solutions of charged deforming AdS black hole, we start with this ans$\ddot{a}$tze of metric,
\begin{subequations}
\begin{multline}
\mathrm{d}s^2=\frac{L^2}{(1-y^2)^2}\Bigg\{-y^2\tilde{\Delta}(y)U_1\mathrm{d}t^2+\frac{4\,y_+^2 U_2\,\mathrm{d}y^2}{\tilde{\Delta}(y)}+y_+^2 \Bigg[\frac{4\,U_3}{2-x^2}\left(\mathrm{d}x+x \sqrt{2-x^2}\,y\,U_4\, \mathrm{d}y\right)^2\\+(1-x^2)^2 U_5\,\left(\mathrm{d}\phi+y^2x\sqrt{2-x^2}\,U_6\,\mathrm{d}t\right)^2 \Bigg]\Bigg\},
\label{eq:ansatzbh}
\end{multline}
\noindent with
\begin{equation}
\Delta(y)=\frac{q^2(1-y^2)^2}{L^2y_+^2}+(1-y^2)^2+y_+^2 (3-3y^2+y^4)\,,\quad\text{and}\quad \tilde{\Delta}(y)= \Delta(y) \delta + y_+^2 (1 - \delta)\,,
\end{equation}\label{key}
\end{subequations}
where the functions $U_i,i\in(1,...,6)$ depend on $x$ and $y$, the parameter $q$ is the charge of black hole, and $y_{+}$ is horizon radius. Here, $y$ is related to radial coordinate $r$ with $r=Ly_+/(1-y^2)$,  and $x$ represents polar angle on $S^2$ with $\sin\theta=1-x^2$. When $U_1=U_2=U_3=U_5=\delta=1$ and $U_4=U_6=0$, the line element (\ref{key}) can reduce to RN-AdS black hole.
\par
Considering an axial symmetry system, we have polar angle  reflection symmetry $\theta\rightarrow\pi-\theta$ on the equatorial plane, and thus it is convenient to consider the
coordinate range $\theta \in [0,\pi/2] $, i.e $x \in [0,1] $. We require  the functions to satisfy the following boundary conditions
on the equatorial plane $x=0$,
\begin{equation}
\partial_x U_i(0,y)=0,  \;\;\;i=1,2,3,4,5,6,
\end{equation}
and  set  axis boundary conditions at $x=1$, where regularity must be imposed Dirichlet boundary conditions on
$U_4$ and Neumann boundary conditions on the other functions
\begin{equation}\label{abc}
U_4(1,y)=0,
\end{equation}
and
\begin{equation}\label{abc}
\partial_x U_1(1,y)=\partial_x U_2(1,y)=\partial_x U_3(1,y)=\partial_x U_5(1,y)=\partial_x U_6(1,y)=0.
\end{equation}
 At $y=1$, we set $U_4=0$, $U_{6}=\varepsilon$ and $U_{1}=U_{2}=U_{3}=U_{5}=1$. Moreover, expanding the equations of motion near  $x=1$  gives $U_3(1,y)=U_5(1,y)$.

In order to ensure that the number of unsolved functions is the same as that of equations in Deturk method, we introduce the component $A_\phi$ in gauge potential. We choose the following form of gauge potential
 \begin{equation}\label{gauge}
   A = A_t dt+ A_\phi d\phi ,
 \end{equation}
 where $A_t$ and $A_\phi$ are all real functions of $x$ and
$y$.   As for the boundary conditions of vector field, we set $A_t(x,1)=\mu$ and $A_x(x,1)=0$, where the constant $\mu$ is chemical potential which represents the asymptotic behavior of Maxwell field at infinity. At $x=1$, we choose $A_t(1,y)=0$ and $A_x(1,y)=0$.
\par
 The Hawking temperature of charged deforming black hole under ans$\ddot{a}$tze (\ref{key}) takes the following form:
  \begin{equation}
  T=\frac{1}{4\pi}\sqrt{-g^{tt}g^{MN}\partial_{M}g_{tt}\partial_{N}g_{tt}}\mid_{r=r_{+}}=\frac{y_{+}^4+\delta(-q^2+y_{+}^2(1+2y_{+}^2))}{4\pi y_{+}^3}.
  \label{temperature}
  \end{equation}
  When the charge $q=0$, the formula (\ref{temperature}) can reduce to the temperature of Schwarzschild-AdS black hole which was also given in \cite{Markeviciute:2017jcp}. When we fix $\delta=1$, the extremums of temperature $T$  depend on the value of charge $q$:
\begin{itemize}
	\item $q=0$: There is only a local minimum $T_{min}=T_S=\frac{\sqrt{3}}{2\pi}$, which is the minimal temperature of Schwarzschild-AdS black hole.
	\item $0<q<1/6$: There are two extremums of temperature.
	\begin{equation}
	\left\{
	\begin{aligned}
	&T_{min}=\frac{3\sqrt{\frac{3}{2}}\left(-q^2+\frac{1}{12}\left(\sqrt{1-36q^2}+1\right)^2+\frac{1}{6}\left(\sqrt{1-36q^2}+1\right)\right)}{\pi\left(\sqrt{1-36q^2}+1\right)^{3/2}},\\
	&T_{max}=\frac{3\sqrt{\frac{3}{2}}\left(-q^2+\frac{1}{12} \left(1-\sqrt{1-36 q^2}\right)^2+\frac{1}{6}\left(1-\sqrt{1-36q^2}\right)\right)}{\pi\left(1-\sqrt{1-36 q^2}\right)^{3/2}}
	.
	\end{aligned}
	\right.
	\end{equation}
	\item $q=1/6$: $T_{max}=T_{min}=T_{RN}=\frac{\sqrt{6}}{3\pi}$, which is the minimal extremum of temperature for RN-AdS black hole.
    \item $q>1/6$: There exists no extremum of temperature.

\end{itemize}
 \begin{figure}[!]
	\centering
	\includegraphics[width=0.8\textwidth]{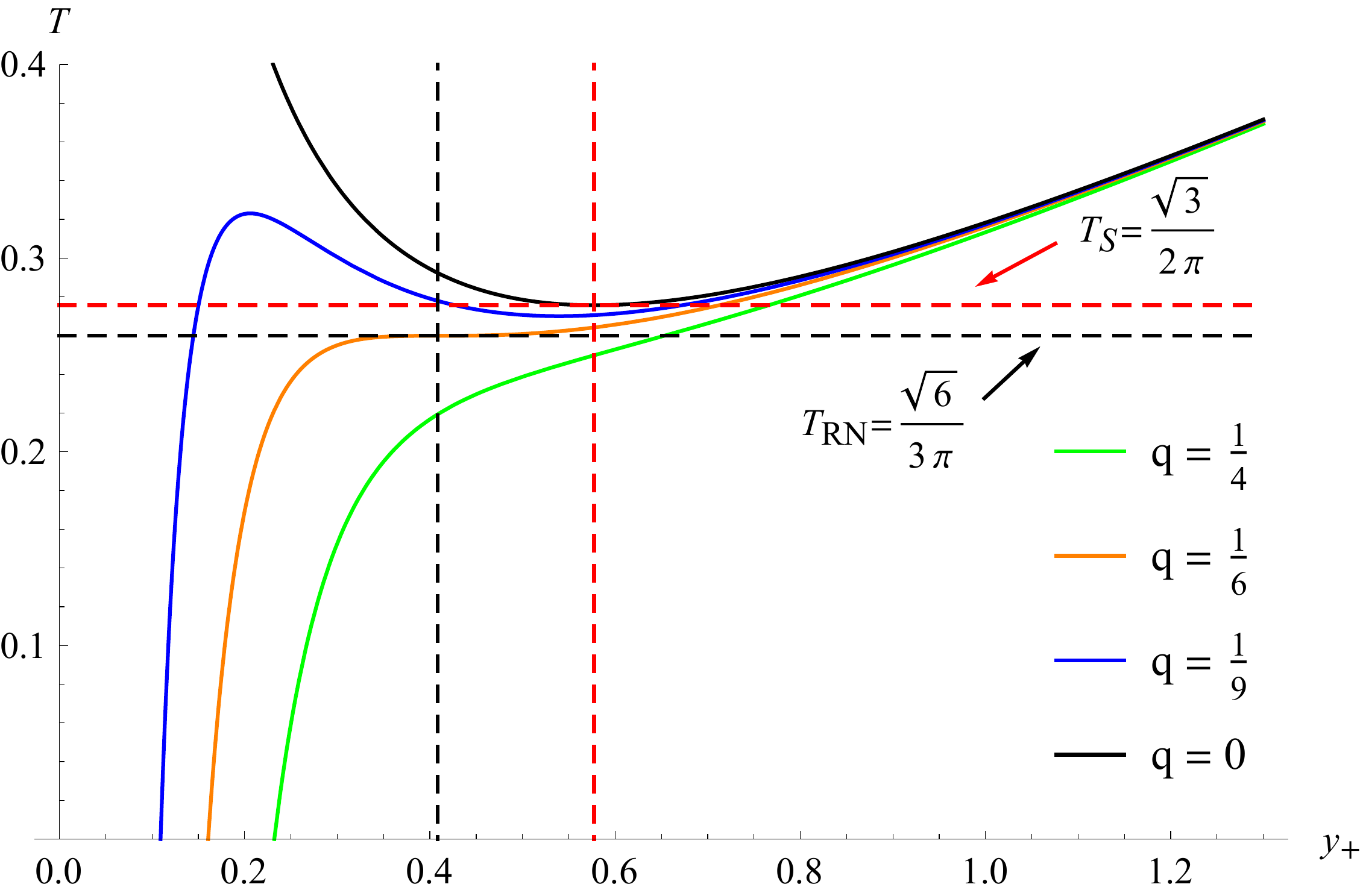}
	\caption{The temperature $T$ as  functions of $y_+$ for $\delta$=1. From top to bottom, the black, blue, red and green lines describe charge $q=0$, $\frac{1}{9}$, $\frac{1}{6}$ and $\frac{1}{4}$, respectively. The red and black horizontal dashed lines represent $T_{S}=\frac{\sqrt{3}}{2\pi}$ and $T_{RN}=\frac{\sqrt{6}}{3\pi}$. The red and black vertical lines represent $y_{+}=\frac{1}{\sqrt{3}}$ and  $y_+=\frac{1}{\sqrt{6}}$.}
	\label{temf}
\end{figure}

\par
Next, we will analyze how the charge $q$ and temperature $T$ determine the number of values of horizon. In Fig.$\ $\ref{temf}, we plot the temperature $T$ as functions of horizon $y_+$ at $\delta=1$ for several values of charge $q$. The black, blue, orange and green lines represent $q=0,\frac{1}{9}, \frac{1}{6}$ and $\frac{1}{4}$, respectively. For $q=\frac{1}{6}$, the intersection of the black horizontal and the vertical dashed lines indicates horizon $y_+=\frac{1}{\sqrt{6}}$ and $T_{RN}=\frac{\sqrt{6}}{3\pi}$. For $q=0$, the intersection of the red horizontal and the vertical dashed lines indicates horizon $y_+=\frac{1}{\sqrt{3}}$ and $T_{S}=\frac{\sqrt{3}}{2\pi}$. The number of values of horizon depends on different ranges of temperature $T$ and charge $q$:

\begin{enumerate}
	\item $q=0$:
		\begin{enumerate}
		\item  $T<T_{S}$: There exists no horizon.
		\item  $T=T_{S}$: There are two equal values of horizon $y_+=\frac{1}{\sqrt{3}}$.
		\item  $T>T_{S}$: There are two different values of horizon.
	\end{enumerate}
	\item $0<q<1/6$:
	\begin{enumerate}
		\item  $T<T_{min}$ or $T>T_{max}$: There is one value of horizon.
		\item  $T=T_{min}$ or $T=T_{max}$: There are three values of horizon and two of them are equal.
		\item  $T_{min}<T<T_{max}$: There are three different values of horizon.
		\end{enumerate}
	\item $q=1/6$:
	\begin{enumerate}
	\item  $T=T_{RN}$: There are three equal values of horizon $y_+=\frac{1}{\sqrt{6}}$.
	\item  $T\neq T_{RN}$: There is only one value of horizon.
\end{enumerate}
	\item $q>1/6$: There is only one value of horizon.
\end{enumerate}

\par
 By regulating parameter $\delta$, we can also get three values of horizon below the local minimal temperature $T_{min}$. For simplify, we fix chemical potential $\mu=0.5$ and AdS radius $L=1$ in our numeral calculations.


\numberwithin{equation}{subsection}	

\par
\begin{figure}[!hbt]
	\centering
	\begin{minipage}[t]{0.48\textwidth}
		\centering
		\includegraphics[width=6.5cm]{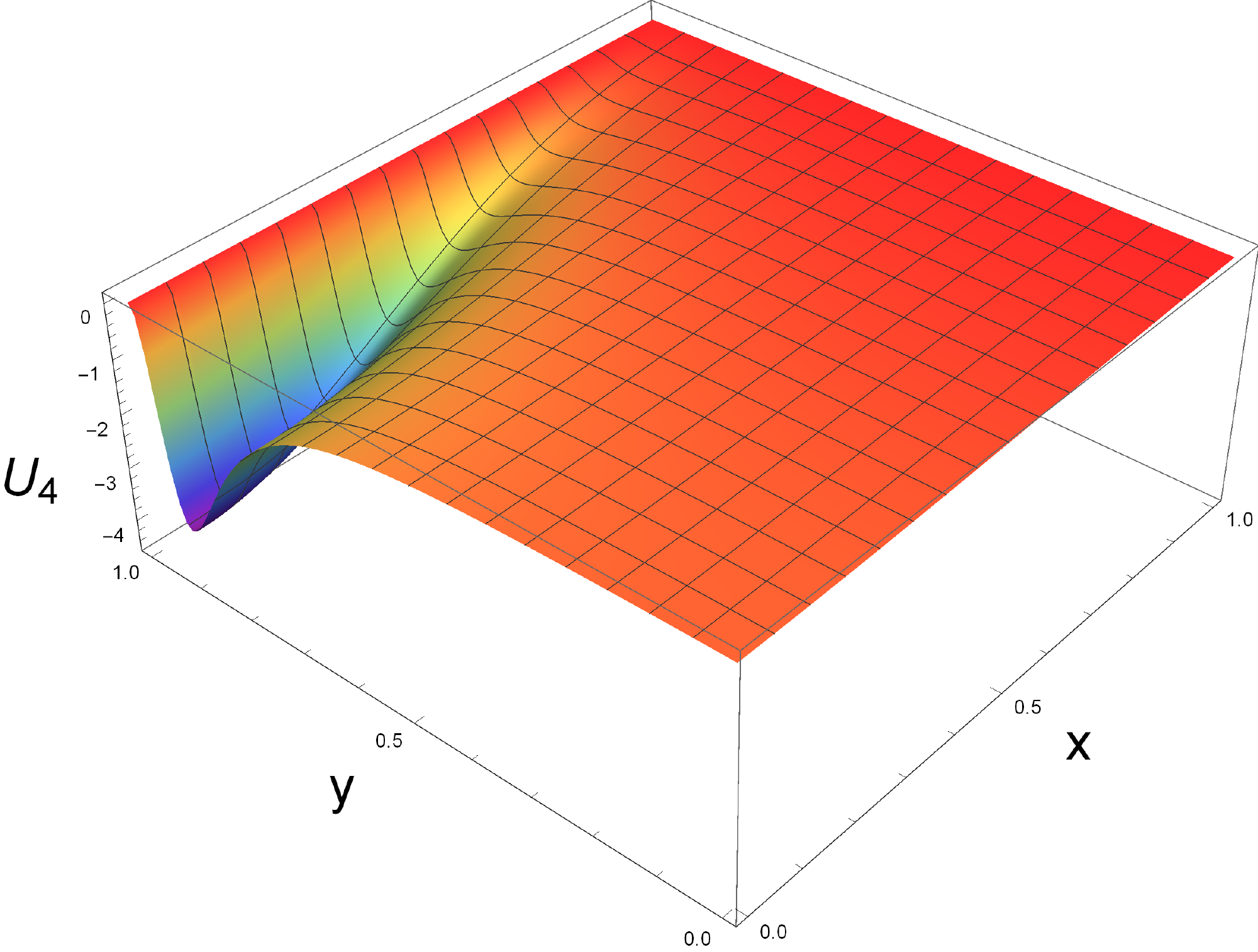}
	\end{minipage}
	\begin{minipage}[t]{0.48\textwidth}
		\centering
		\includegraphics[width=6.5cm]{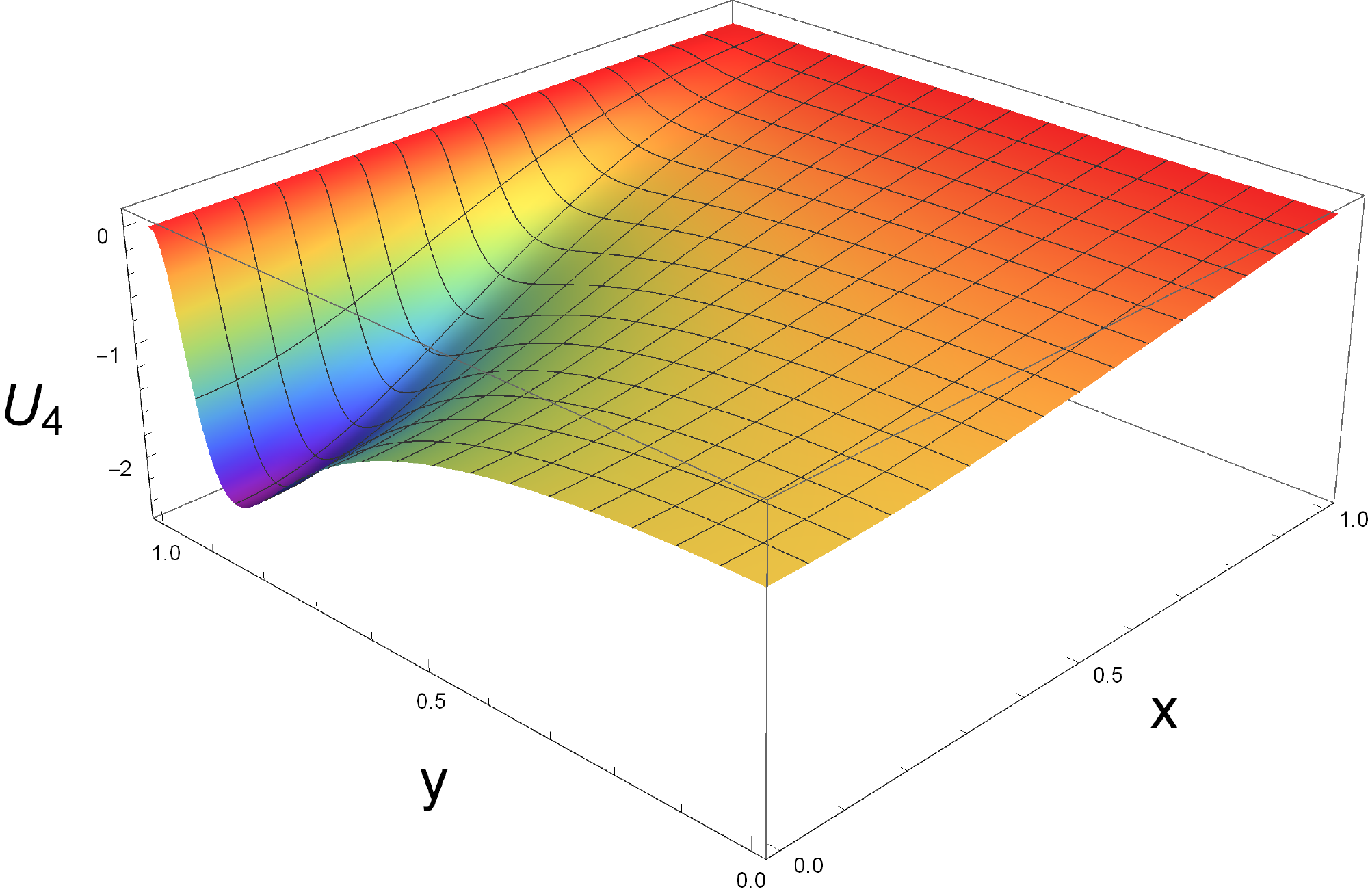}
		\end{minipage}
\begin{minipage}[t]{0.48\textwidth}
		\centering
		\includegraphics[width=6.5cm]{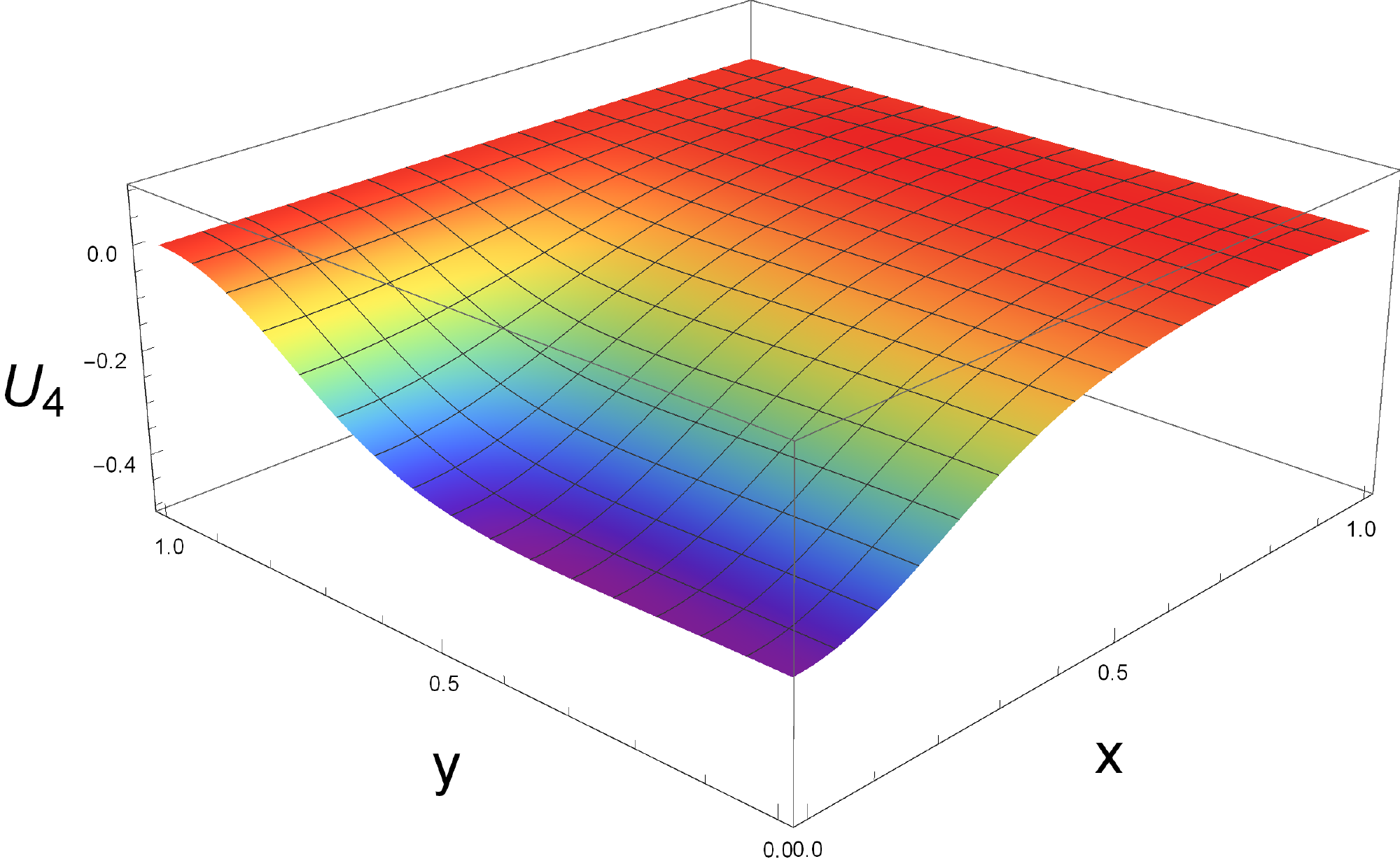}
		\end{minipage}
\begin{minipage}[t]{0.48\textwidth}
		\centering
		\includegraphics[width=6.5cm]{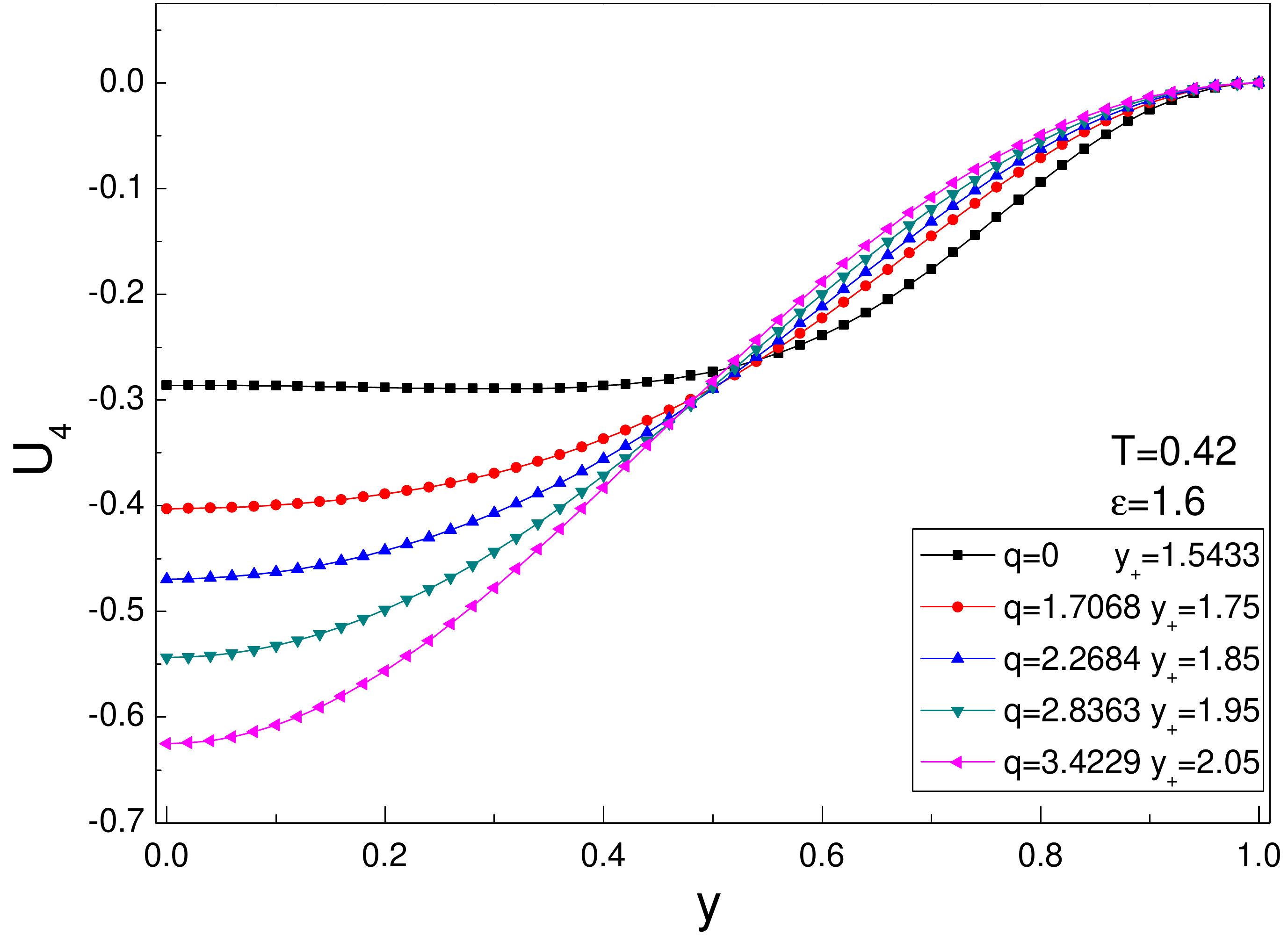}
		\end{minipage}
	\caption{\emph{Top}: The distributions of $U_4$ as functions of $x$ and $y$  of two small branches for $y_+=0.0992$ (left) and $y_+=0.1773$ (right). \emph{Bottom left}: The distributions of $U_4$ as functions of $x$ and $y$  for large branch $y_+=1.5436$. The three solutions of $U_4$ are given with $T=0.42$, $\varepsilon=1.6$, $\delta=1$ and $q=0.07057$.  \emph{Bottom right}: The distributions of $U_4$ as functions of $y$ at the equatorial plane $x=1$ for several values of charge $q$ with $T=0.42$ and $\varepsilon=1.6$. From top to bottom, the black, red, blue, green and pink lines describe the charge $q=0$, $1.7068$, $2.2684$, $2.8363$ and $3.4299$, respectively.}
\label{u}	
\end{figure}

 In Fig.$\ $\ref{u}, we give the typical distributions of $U_4$ as functions of $x$ and $y$ for $T=0.42$, $\varepsilon=1.6$ and $\delta=1$. When fixing $q=0.07057<\frac{1}{6}$, we can obtain three values of horizon.  The distributions of $U_4$  for two small branches with $y_+=0.0992$ (left) and $y_+=0.1773$ (right) are given in the top of Fig.$\ $\ref{u}. The left of bottom is the distributions of $U_4$ for large branch $y_+=1.5436$. To understand how the charge $q$ influences the distributions of $U_4$, we also plot $U_4$ as functions of $y$ at the equatorial plane $x=1$ for several values of $q$. From top to bottom, the distributions of function $U_4$ with charge $q=$ $0$, $1.7068$, $2.2684$, $3.4299$ are represented by black, red, blue, green and pink lines, respectively. Due to the existence of relation (\ref{temperature}), the horizon radius $y_+$ increases with the increasing of charge $q$ for a fixed temperature.

\subsection{Horizon geometry}\label{31}

 \hspace*{0.6cm}In this subsection, we will study how the black hole horizon geometry behaves with the increase of boundary rotation parameters $\varepsilon$ and charge $q$. We could use an isometric embedding in the three-dimensional space \cite{c1,c2,c3,c4,c5} to investigate the horizon geometry of a two-dimensional surface in a curved space \cite{Markeviciute:2017jcp,Gibbons:2009qe}. With the method provided by \cite{Markeviciute:2017jcp}, the horizon of black hole is embedded into hyperbolic $H^3$ space in global coordinates:
  \begin{equation}
  ds^2_{H^3}=\frac{dR^2}{1+R^2/l^2}+R^2\left[\frac{dX^2}{1-X^2}+(1-X^2)d\phi^2\right],
  \end{equation}
  where $l$ is the radius of the hyperbolic space and we fix $l=0.73$ in our whole calculation. The induce metric of the horizon of black hole is the following form:
  \begin{equation}
  ds^2_{H}=L^2\left[\frac{4y_{+}^2U_{3}(x,0)}{2-x^2}dx^2+y_{+}^2(1-x^2)^2U_{5}(x,0)d\phi^2\right],
  \label{reduce1}
  \end{equation}
  which can be obtained from  the ans$\ddot{a}$tze (\ref{key}). The embedding is given by a curve with two parameters $\{R(x),X(x)\}$ and written by:
  \begin{equation}
  ds_{pb}^2=\left[\frac{R(x)'}{1-\frac{R(x)^2}{l^2}}+\frac{R(x)'^2X(x)'^2}{1-X(x)^2}\right]dx^2+R(x)^2(1-X(x)^2)d\phi^2.
  \label{reduce2}
\end{equation}
Equating this line element with induce metric (\ref{reduce1}), we can get the following first order differential equation:
\begin{eqnarray}
0&=&4H(x)P(x)(X(x)^2-1)^2[P(x)-l^2(X(x)^2-1)]  \\
&&+4l^2P(x)X(x)(X(x)^2-1)P(x)'X(x)'-(X(x)^2-1)^2l^2P(x)^2(l^2+P(x))^2X(x)'^2, \nonumber
\end{eqnarray}
where $H(x)=(2-x^2)^{-1}(4y_+^2U_3(X,0))$ and $P(X)=y_+^2(1-x^2)^2U_5(x,0)$.
\begin{figure}[hhh]
	\centering
	\includegraphics[width=0.9\textwidth]{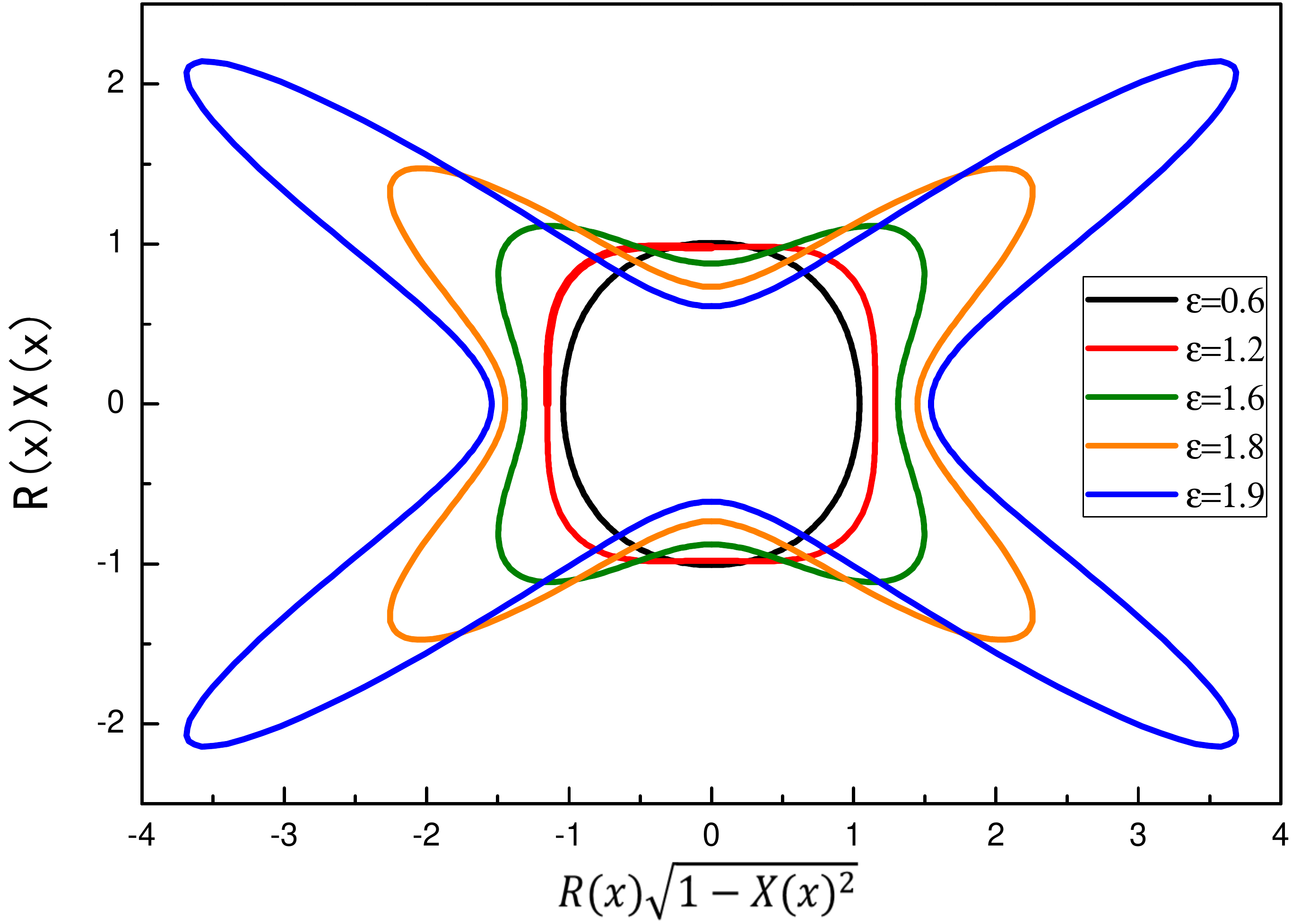}
	\caption{Hyperbolic embedding of the cross section of the large black hole horizon for different values of $\varepsilon$ with $T=0.42$ and $q=0.07057$.}
	\label{largehorizon}
\end{figure}
\par
 In Fig.$\ $\ref{largehorizon}, we show the hyperbolic embedding of the cross section of the large black hole horizon  for different values of $\varepsilon$ with the charge $q=0.07057$ and the temperature $T=0.42\geq T_{min}=0.2325$. From inner to outer, the black, red, green, orange and blue lines describe the boundary rotation parameter  $\varepsilon$=$0.6$, $1.2$ ,$1.6$, $1.8$ and  $1.9$, respectively. It is clear that the horizon deforms more dramatically with the increase of boundary rotation parameter $\varepsilon$.

\begin{figure}[t]
	\centering
	\begin{minipage}[c]{0.5\textwidth}
		\centering
		\includegraphics[scale=0.26]{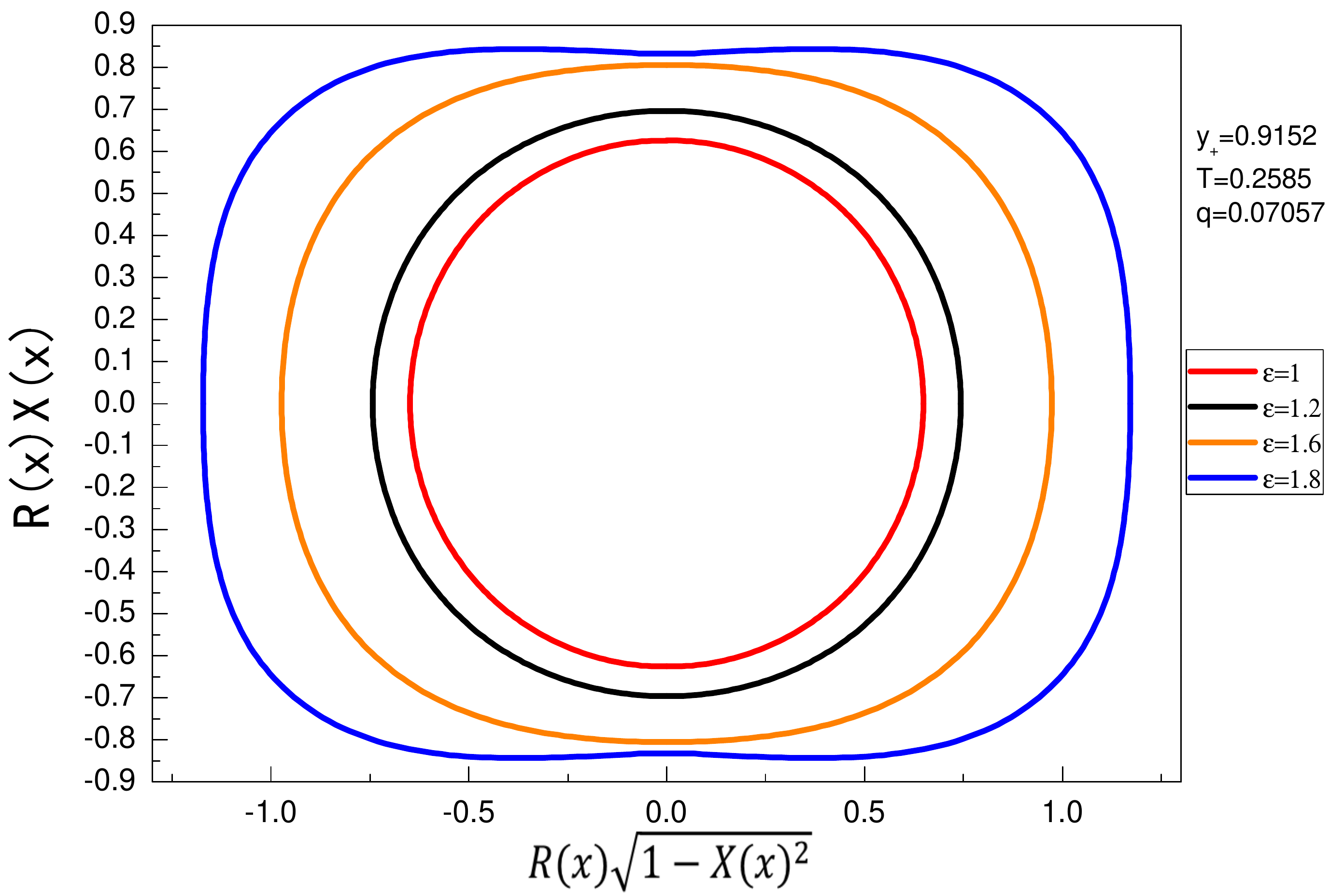}
	\end{minipage}%
	\begin{minipage}[c]{0.5\textwidth}
		\centering
		\includegraphics[scale=0.26]{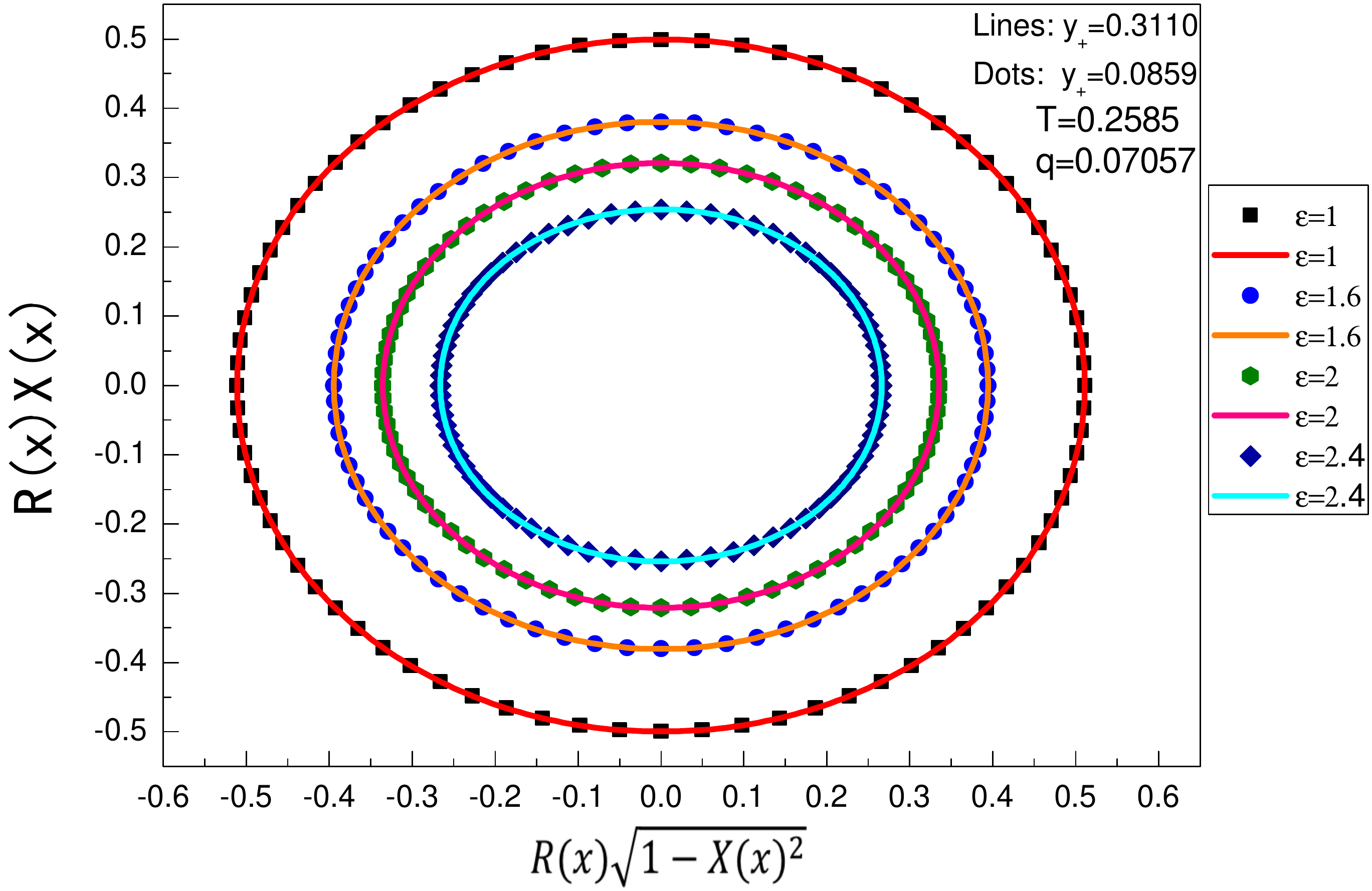}
	\end{minipage}
	\caption{Hyperbolic embedding of the cross section of three black hole horizons  for different values of $\varepsilon$ at $T=0.2585$ and $q=0.07057$. \emph{Left}: The horizon geometry of large branch for $y_+=0.9152$. \emph{Right}: The horizon geometry of two small branches for $y_+=0.3110$(lines) and $y_+=0.0859$(dots).}
	\label{fig4}
\end{figure}
\par
Considering there is only one horizon radius when $T<T_{min}$ with $\delta=1$,  we adjust $\delta<1$ to get three values of horizon and study the deformation of horizon for a fixed low temperature. In Fig.$\ $\ref{fig4}, we present hyperbolic embedding of the cross section of three black hole horizons for different values of $\varepsilon$ with $T=0.2585$ and $q=0.07057$. In the left panel, we show large black hole solutions for $y_+=0.9152$ and find that the size of
the deformation of horizon cross section  increases as $\varepsilon$ increases, which is similar to the situation in Fig.$\ $\ref{largehorizon}. In the right panel, we show the result of two small branches for $y_+=0.3110$(lines) and $y_+=0.0859$(dots). What is different from the left panel is  that the size of the deformation of horizon is a decreasing function of boundary rotation parameters $\varepsilon$. For the two small branches with different horizon radii, the horizon radius of the bigger one is nearly four times as that of the smaller one, but it is interesting to find that the two small branches have same embedding graphs of horizon geometry.

\begin{figure}[hhh]
	\centering
	\includegraphics[width=0.8\textwidth]{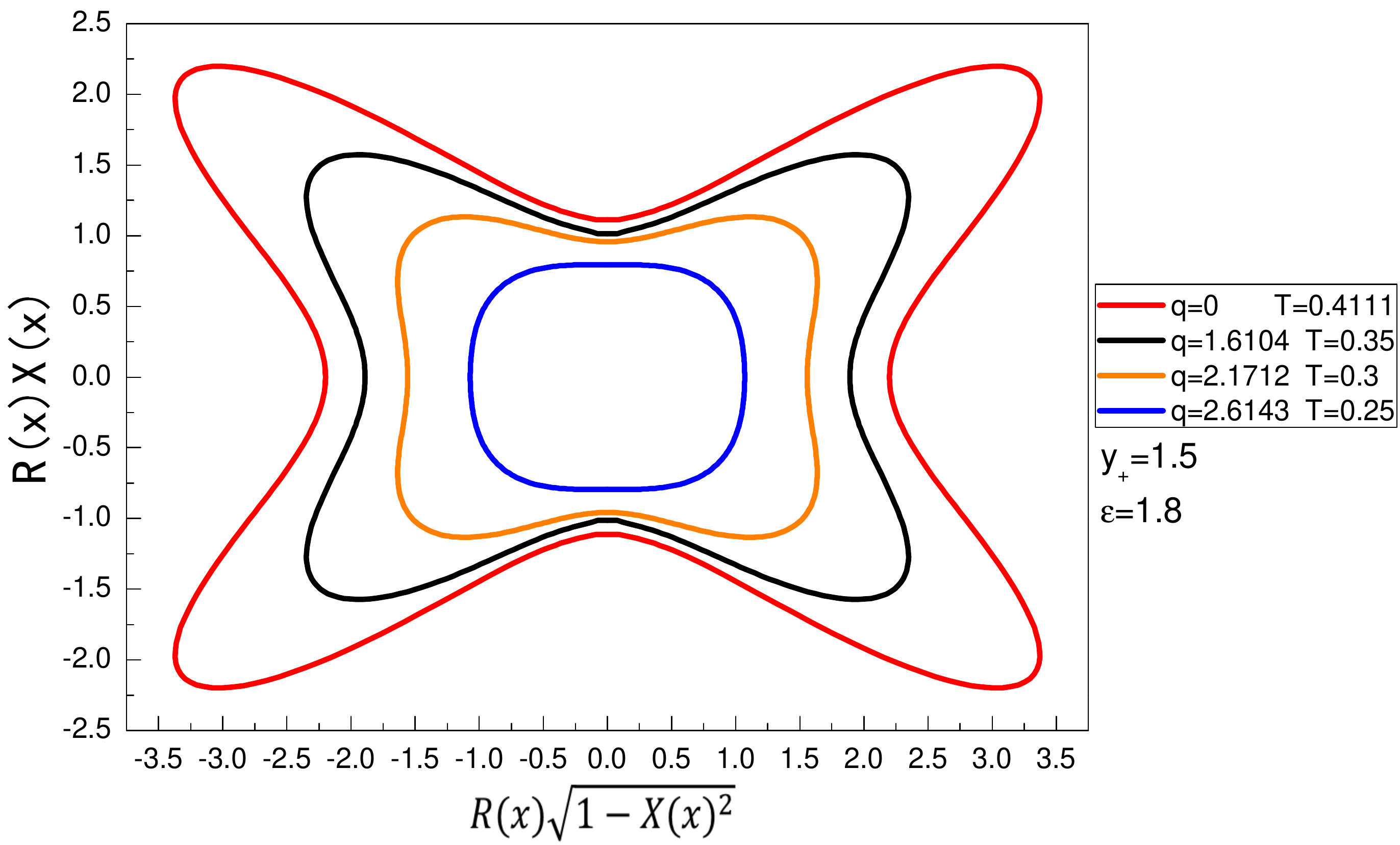}
	\caption{Hyperbolic embedding of the cross section of the large black hole horizon  for different values of charge $q$ with $y_+=1.5$ and $\varepsilon=1.6$. }
	\label{fig5}
\end{figure}
\par
To show the effect of charge $q$ on the deformation of horizon, we give hyperbolic embedding of the cross section of the large black hole horizon  for different values of charge $q$ with $y_+=1.5$ and $\varepsilon=1.6$ in Fig.$\ $\ref{fig5}. Due to the existence of relation (\ref{temperature}), the temperature decreases with the increasing of charge $q$ for a fixed horizon radius. From outer to inner, the red, black, orange and blue lines represent $q=0$, $1.6104$, $2.1712$ and $2.6143$ respectively. The deformation of horizon becomes smaller as the charge $q$ increases.

 \subsection{Entropy}

  \hspace*{0.6cm}In this subsection, we will discuss the entropy of  deforming charged black hole. The formula of entropy of black hole is written as
 \begin{equation}
 S=\frac{A}{4G_N}=\frac{2\pi y_+^2L^2}{G_N}\int^1_0dx\frac{1-x^2}{\sqrt{2-x^2}}\sqrt{U_3(x,0)U_5(x,0)}.
 \end{equation}
 \par
\begin{figure}[t]
	\centering
	\begin{minipage}[c]{0.5\textwidth}
		\centering
		\includegraphics[scale=0.27]{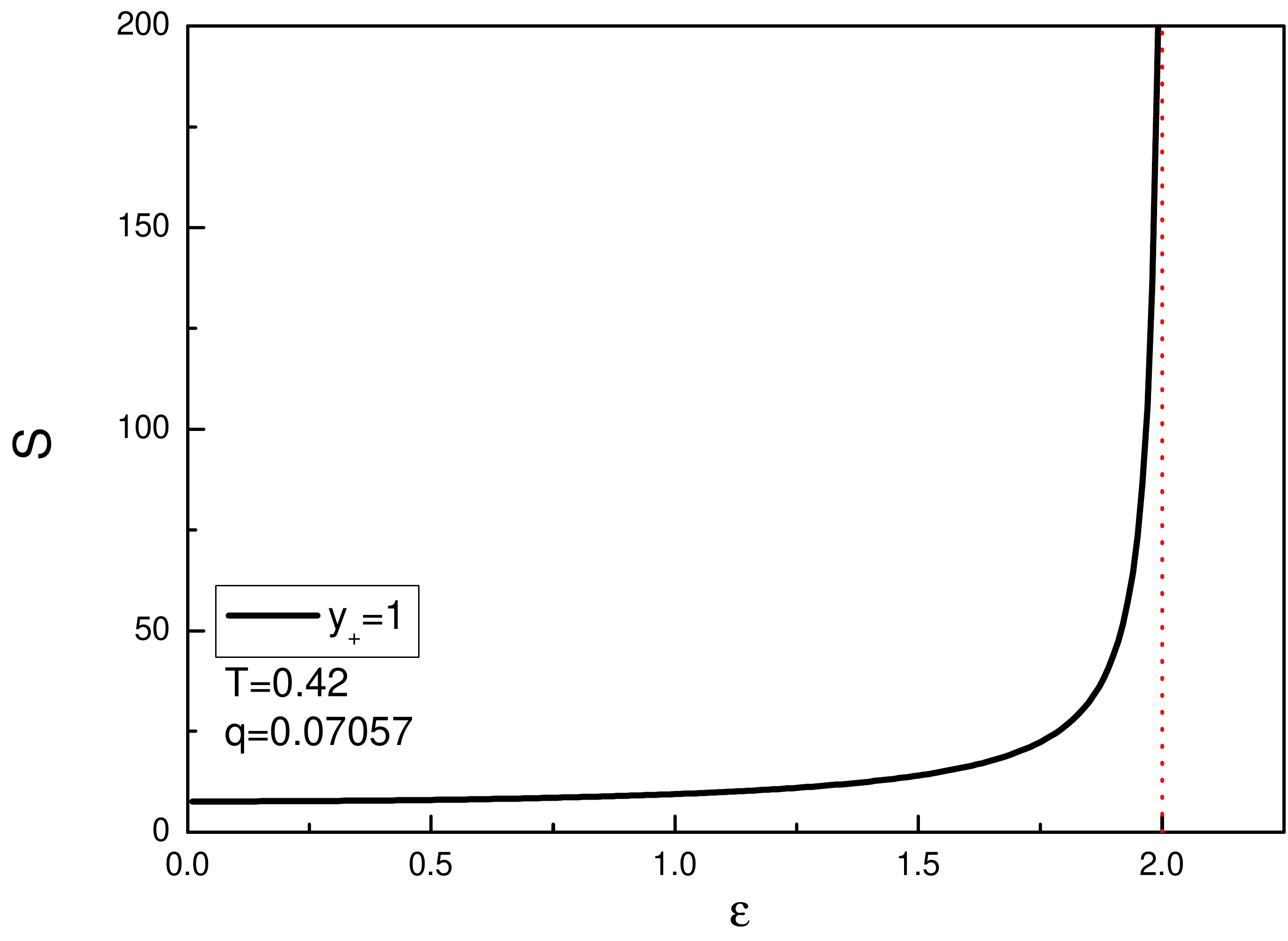}
	\end{minipage}%
	\begin{minipage}[c]{0.5\textwidth}
		\centering
		\includegraphics[scale=0.25]{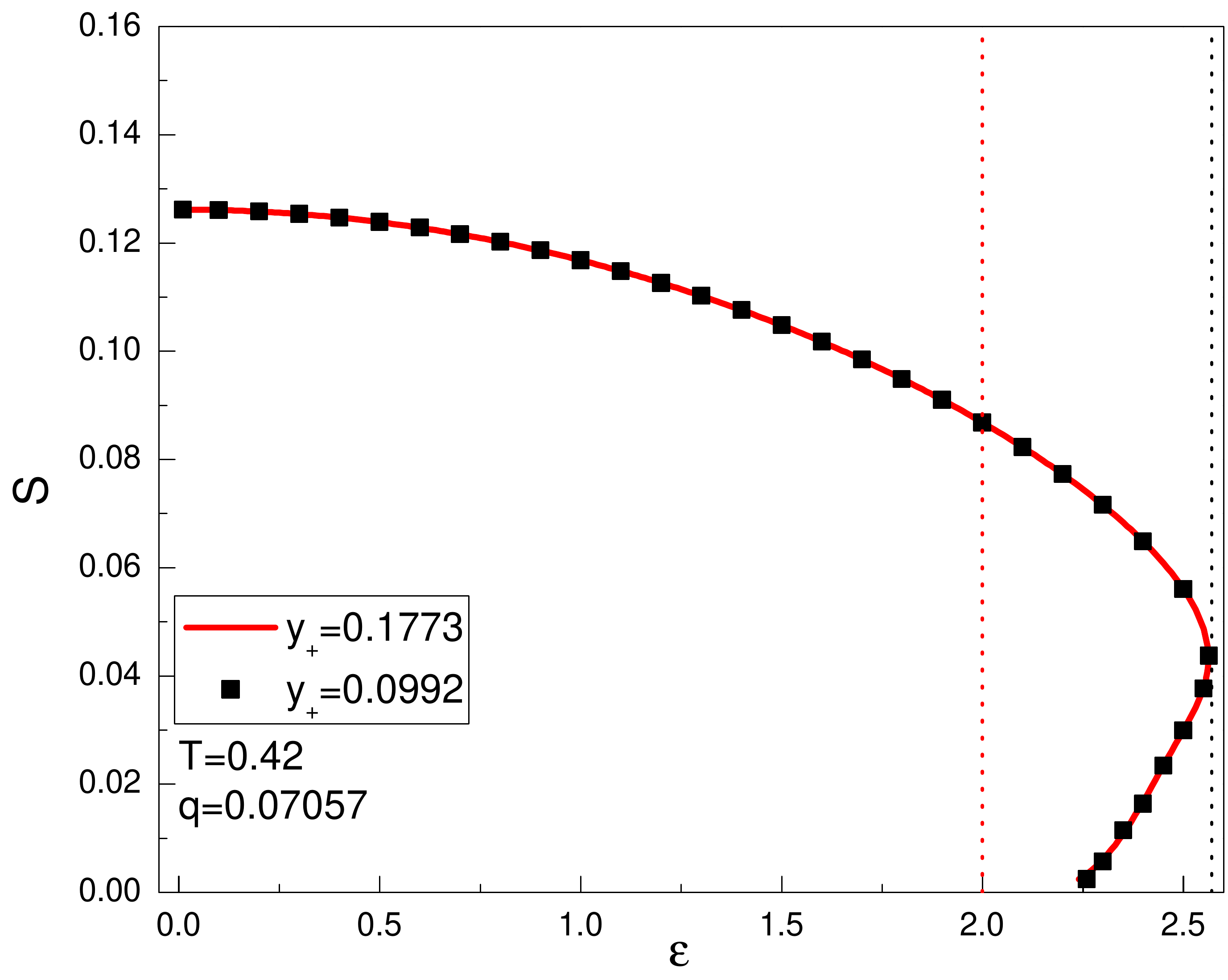}
	\end{minipage}
	\caption{The entropy as functions of boundary rotation parameter $\varepsilon$ for the temperature $T=0.42$ and the charge $q=0.07057$. \emph{Left}: The entropy against boundary rotation parameter $\varepsilon$ for the large branch of black hole solutions $y_+=1$. \emph{Right}: The entropy against boundary rotation parameter $\varepsilon$ for two small branches of black hole $y_+=0.1773$(black dots) and $y_{+}=0.0992$(red line). The vertical red dot lines represent $\varepsilon=2$.}
	\label{42entropy}
\end{figure}

In Fig.$\ $\ref{42entropy}, we show  the entropy against boundary rotation  parameter $\varepsilon$ with $T=0.42$ and $q=0.07057$. The large black hole with $y_+=1$ is shown in the left panel,  while in the right panel, two small branches with $y_+=0.1773$ and $y_+=0.0992$ are represented by red line and black dots respectively. For the large black hole, the entropy is a increasing function of boundary rotation  parameter $\varepsilon$. The entropy approaches infinity as $\varepsilon\rightarrow2$, and we could not find charged deforming black hole solutions when $\varepsilon>2$. As for two small branches with a fixed temperature, the entropy decreases with the increase of boundary rotation parameter $\varepsilon$, and there exist solutions when  $\varepsilon>2$. Furthermore, we also find another family of small black hole solutions, and in these solutions, the entropy increases with the increase of $\varepsilon$.
\par

\begin{figure}[t]
	\centering
	\begin{minipage}[c]{0.5\textwidth}
		\centering
		\includegraphics[scale=0.27]{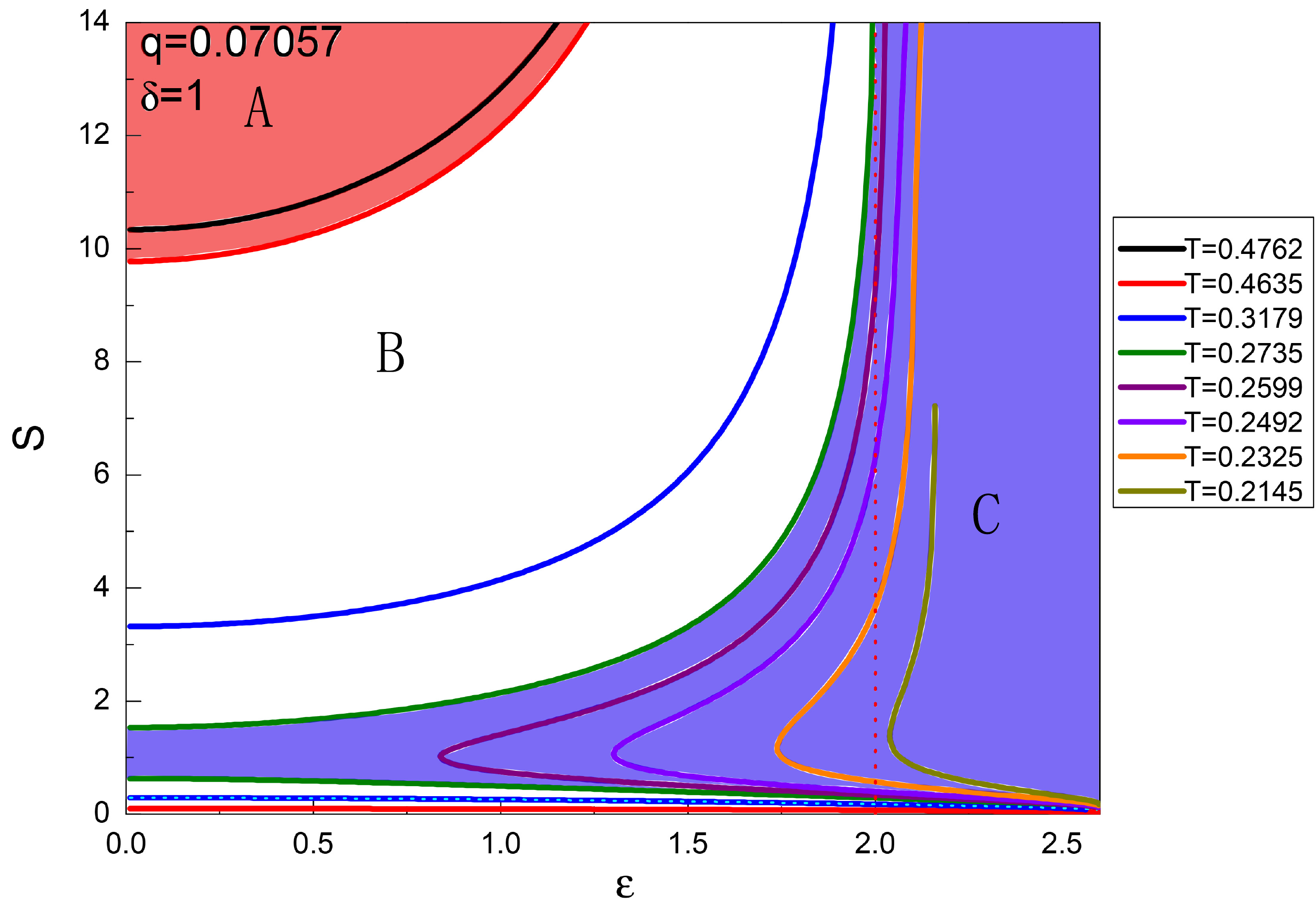}
	\end{minipage}%
	\begin{minipage}[c]{0.5\textwidth}
		\centering
		\includegraphics[scale=0.24]{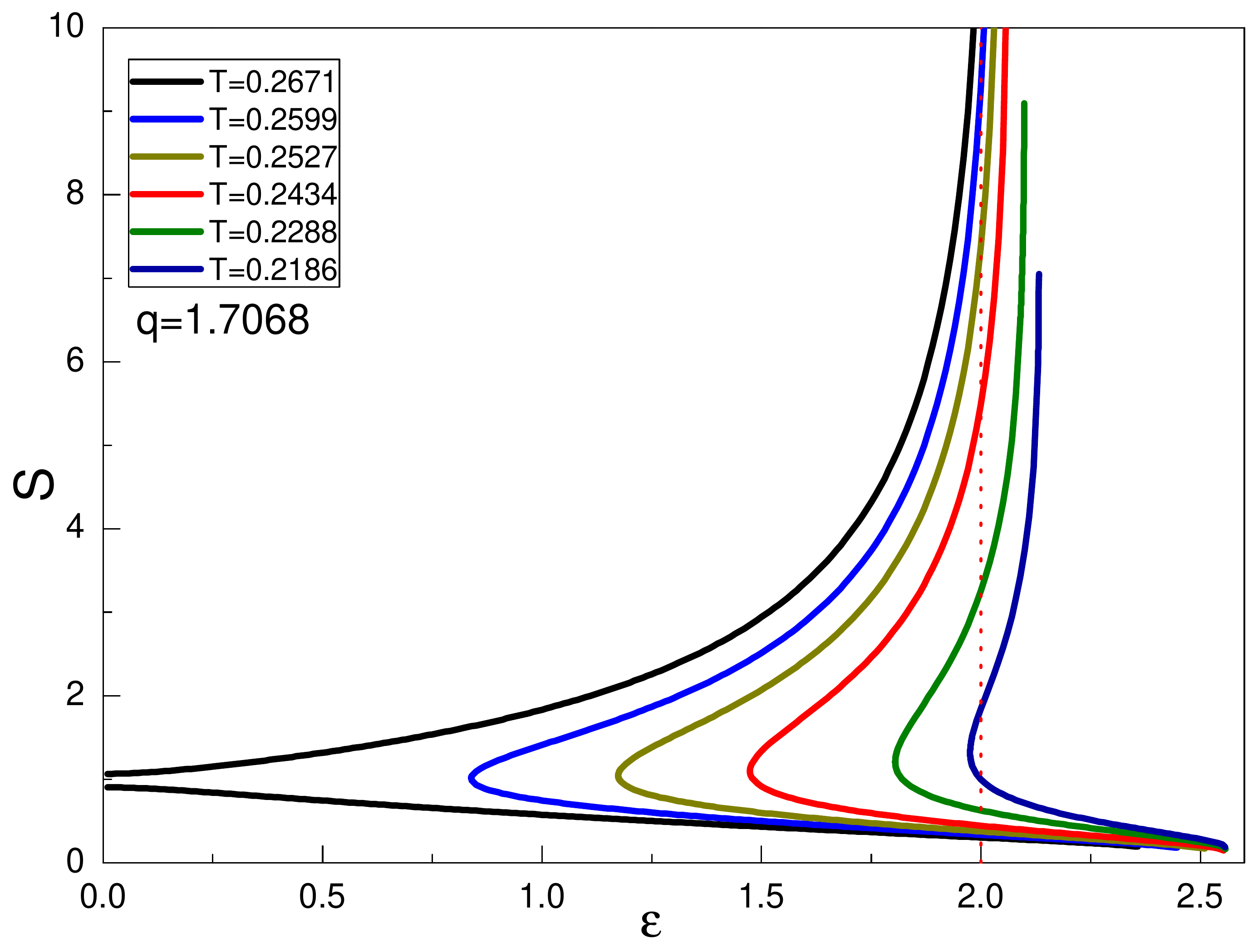}
	\end{minipage}
	\caption{The entropy as functions of boundary rotation parameter $\varepsilon$ for different values of temperature $T$ with $\delta=1$. \emph{Left}: For $q=0.07057$, local maximum and minimum of temperature $T$ are equal to $0.4635$ and $0.2735$ represented by red and green lines respectively. The two extremums divide the phase diagram of entropy into three regions. \emph{Right}: For $q=1.7068$, there is always one solution without extremum of temperature $T$. The vertical red dot lines represent $\varepsilon=2$. }
	\label{deltaentropy}
\end{figure}

To obtain the complete phase diagram of entropy for $\delta=1$, we investigate the whole region of temperature in terms of entropy. We show the entropy as functions of boundary rotation parameter $\varepsilon$  for different values of temperature $T$ with $\delta=1$ in Fig.$\ $\ref{deltaentropy}. In the left panel, when we fix $q=0.07057$, there are two local extremums $T_{max}=0.4635$ and $T_{min}=0.2735$,  the entropy of which are represented by red and green lines respectively. The two extremums divide the temperature into three regions:

\begin{itemize}
	\item Region A with $T>T_{max}$: There is only one value of horizon for a fixed temperature and the entropy increases with the increasing of boundary rotation parameter $\varepsilon$. The region A is indicated by the red area.
	\item Region B with $T_{min}<T<T_{max}$: There are three values of horizon for a fixed temperature. For the large branch of black hole, the entropy increases with $\varepsilon$. Although these two small branches have different black hole horizons, they have same entropy which is a decreasing function of boundary rotation parameter $\varepsilon$.
	\item Region C with $T<T_{min}$: There is only one horizon, but we could find two branches of entropy. The entropy increases with rotation parameter $\varepsilon$ at one branch, while it is a decreasing function of $\varepsilon$ in another branch. It is notable that when $T\leq T_{RN}\approx0.2599$, the two branches of entropy for one temperature join up. The region C is indicated by the blue area.
\end{itemize}
\par
 In the right panel of Fig.$\ $\ref{deltaentropy}, we fix $q=1.7068>\frac{1}{6}$. There only exist one value of horizon for any temperature, but we could obtain two branches of entropy. The entropy increases with  rotation parameter $\varepsilon$ at one branch, while it is a decreasing function of $\varepsilon$ in another branch. It is notable that when $T\leq T_{RN}$, the two branches of entropy could connect, which is similar to the region C in left panel.

\par
\begin{figure}[hhh]
	\centering
	\includegraphics[width=0.9\textwidth]{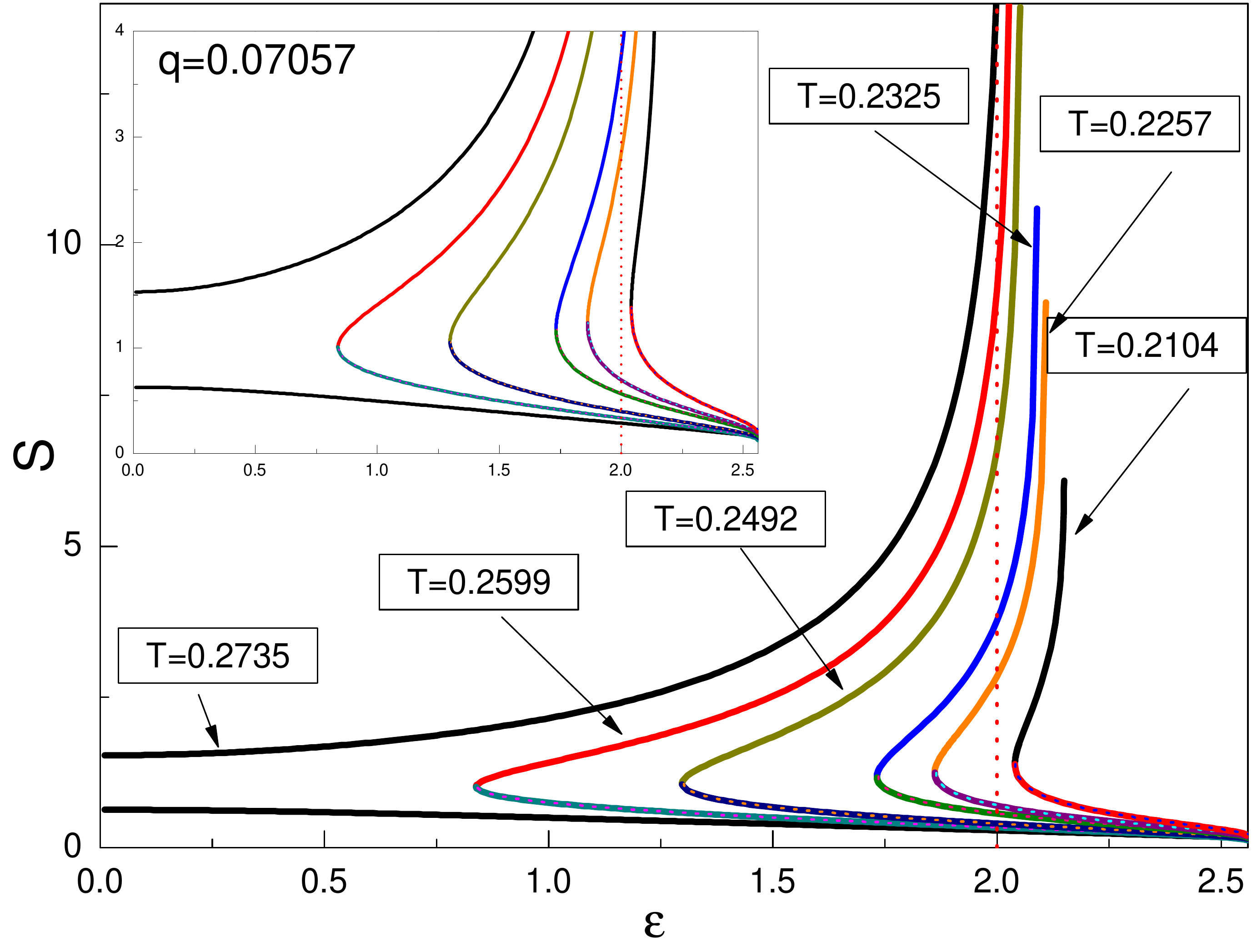}
	\caption{The entropy as functions of boundary rotation parameter $\varepsilon$  for $T\leq T_{min}=0.2735$ with $q=0.07057$. The Black lines indicates $T=T_{min}$ and the red vertical dash line indicates the $\varepsilon=2$.}
	\label{entropy}
\end{figure}

Similar to Subsection \ref{31}, we adjust $\delta<1$ to get three values of horizon and study the entropy for a fixed low temperature. In Fig.$\ $\ref{entropy}, we exhibit entropy as functions of boundary rotation parameter $\varepsilon$  for different values of temperature $T$ at $q=0.07057$. The minimal temperature $T=0.2735$ for $\delta=1$ is represented by black lines. For a fixed temperature $T\leq T_{min}\approx0.2599$,  the large branch join up with two small branches, which form a set of lines. At each set, the line showing that entropy increases with the increasing of $\varepsilon$ describes large branch and corresponding solid line and dot line describe two small branches. From left to right, these sets of lines
indicate $T= 0.2599, 0.2492, 0.2325,0.2257$ and $0.2104$, respectively. Similar to above results in the right panel of Fig.$\ $\ref{42entropy}, the two small branches have same entropy. When temperature is lower than $T_{min}$, the large branches also have solutions with $\varepsilon>2$.  The entropy become infinity when $\varepsilon$ approaches to a maximum value of solutions.
\subsection{Stability}
 \hspace*{0.6cm}In this subsection, we study the stability of deforming charged black hole solutions. Following the method provided in \cite{Markeviciute:2017jcp,d1,d2}, we consider a free, massless and neutral scalar field perturbation to background and solve the Klein-Gordon equation
\begin{equation}
\square\Phi=0 ,
\label{KG}
\end{equation}
and we could decompose the scalar field as the following standard form
\begin{equation}
\Phi=\hat{\Phi}_{\omega,m}(x,y)e^{-i\omega t+im\phi},
\end{equation}
where the constant $\omega$ is the frequency of the complex scalar field and $m$ is the azimuthal harmonic index. Considering the ingoing Eddington-Finkelstein coordinates \cite{d2,Berti:2009kk}, the scalar field with the ans$\ddot{a}$tze of the black hole metric (\ref{key}) could be decomposed into

\begin{equation}
\Phi(t,x,y,\phi)=e^{-i\omega t}e^{i\omega \phi}y^{-i\frac{2\omega y_+}{1+3y_+^2}}(1-y^2)^3(1-x^2)^{|m|}\psi(x,y) ,
\end{equation}
where the powers of $x$ and $y$ are chosen to make function $\psi(x,y)$ regular at the origin. The boundary conditions are imposed as follow:
\begin{equation}
\left\{
\begin{aligned}
&\partial_{x}\psi(x,y)=0, \quad  x=\pm1,\\
&\partial_{y}\psi(x,y)=0, \quad y=0,\\
&-2iy_+\omega \psi(x,y)+(1+3y_+^2)\partial_y\psi(x,y)=0, \quad y=1.
\end{aligned}
\right.
\end{equation}

\begin{figure}[!hbt]
	\centering
	\includegraphics[width=0.8\textwidth]{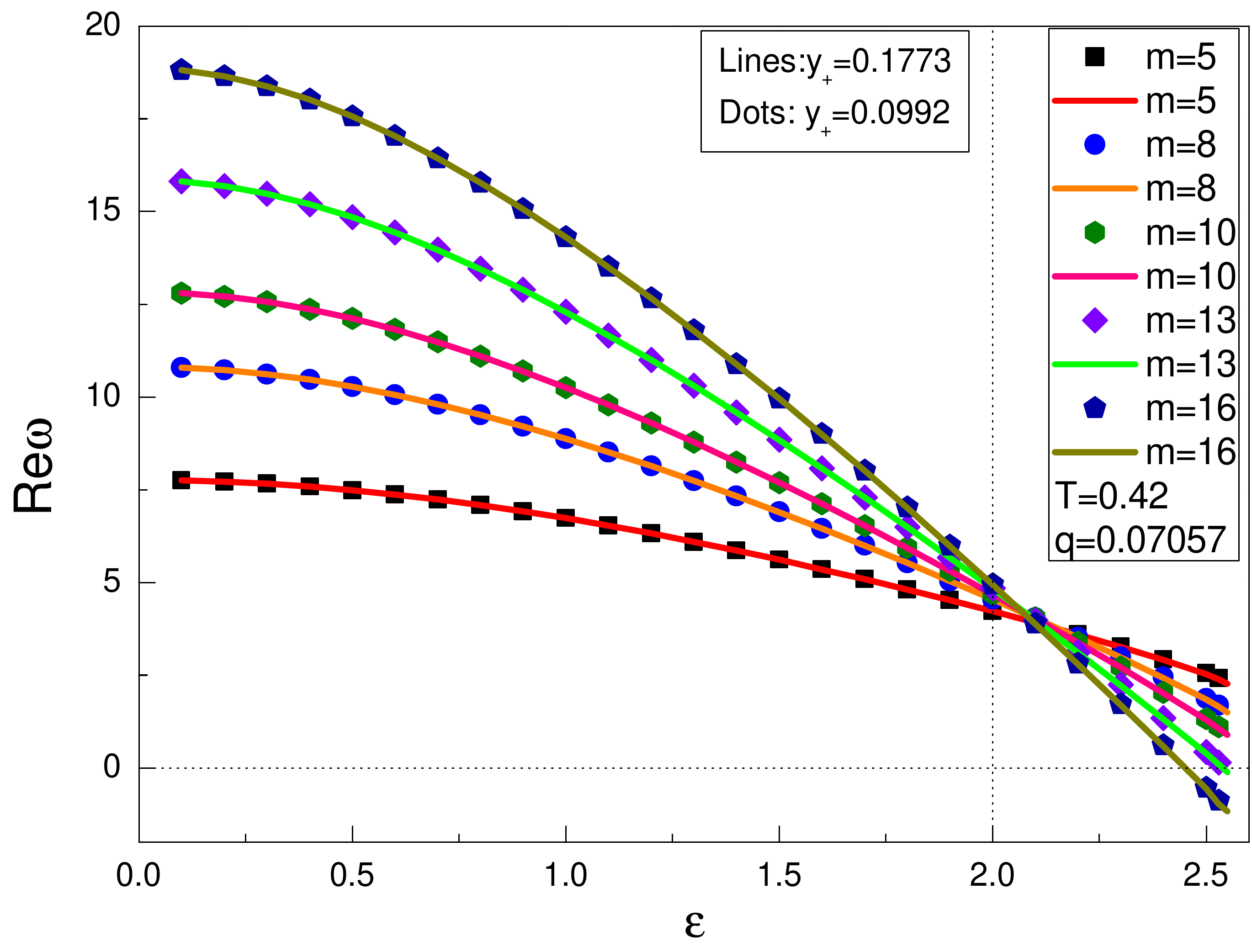}
	\caption{ The real part of frequencies $\omega$ against the rotation parameter $\varepsilon$ of two  small branches for different value of angular quantum number $m$ at $T=0.42$ and $q=0.07057$. The black horizontal line represents $Re$ $\omega=0$. The black vertical horizontal line represents $\varepsilon=2$.}
	\label{QN}
\end{figure}
In Fig.$\ $\ref{QN}, we give the real part of quasinormal frequencies $\omega$ against the rotation parameter $\varepsilon$ of two  small branches for different values of angular quantum number $m$. From top to bottom, these dot lines represent $m=5, 8, 10, 13$ and $16$ respectively. Similar to the above results of horizon geometric and entropy, these two small branches have equal quasinormal frequencies though the horizon radius of the bigger one is nearly twice as that of the smaller one. The real part of frequencies $Re$ $\omega$ is always positive when $m<13$. When $m\geq13$,  $Re$ $\omega$ would appear a negative value with the increase of rotation parameter $\varepsilon$, which means we could obtain a stable deforming charged black hole solution with scalar condensation.

\section{Conclusions and Outlook}\label{Sec4}
 \hspace*{0.6cm}In this paper, we studied the conformal boundary of four-dimensional static asymptotically AdS solutions in Einstein-Maxwell gravity and constructed solutions of deforming charged AdS black hole. In contrast with the situations without charge, the charge $q$ could influence the extremums of temperature $T$ which divide the range of temperature into different regions  according to the value of charge $q$. The number of horizons depends on the different regions of temperature $T$. Moreover, there exists no horizon when $T<T_{min}$ for $q=0$, but when we take charge $q\neq0$, there is at least one value of horizon for a fixed temperature.
 \par
 We also investigated physical properties for charged deforming AdS black holes, including the deformation of horizon, entropy and stability:
 \begin{itemize}
 	\item Deformation of horizon: In the region with three values of horizon for a fixed temperature, the deformation of horizon for large branch increases with the increasing of boundary rotation parameter $\varepsilon$, while that of small branches is a decreasing function of $\varepsilon$, which shows very similar results to the cases without charge. We also studied how the horizon deforms against the charge $q$ and found that the deformation of horizon became smaller as the charge $q$ increases.
 	\item Entropy: In the region with three values of horizon for a fixed temperature,  with the increase of $\varepsilon$, the entropy of large branches increases,  while that of small branches decreases. There  also exist  another set of unstable solutions of small branches, where the entropy increases with the increasing of $\varepsilon$. The entropy of large branch and small branches for a fixed temperature join up when temperature $T$ is lower than $T_{RN}$. It is worth noting that in the region with one value of horizon for a fixed temperature, we could find two families of solutions with same horizon radius, and they have different properties of entropy when the temperature $T<T_{RN}$.
 	\item Stability: We have studied the stability of scalar fields in the background of deforming charged AdS black holes, and found that when angular quantum number $m\geq13$, the real part of frequencies begins to appear negative values, which means scalar condensation.
 \end{itemize}
\par
The most interesting finding in our research is that in the region with three values of horizon at one temperature, the two small branches for a fixed temperature have same numerical results, including deformation of horizon, entropy and stability though their horizon radii might vary many times.

 \par
At present, we have studied the horizon geometry, entropy and stability of charged AdS black hole with  differential rotation boundary. But the angular momentum, energy densities and thermodynamic relation of deforming charged black hole have not been studied, and we hope to
investigate these in our future work. Besides, we are planning to study the deforming charged black holes in $f(R)$ gravity and nonlinear electrodynamics.

\section*{Acknowledgement}
We would like to thank  Yu-Xiao Liu and Jie Yang  for  helpful discussion. Some
computations were performed on the   Shared Memory system at  Institute of Computational Physics and Complex Systems in Lanzhou University. This work was supported by the Fundamental Research Funds for the Central Universities (Grants No. lzujbky-2017-182).

\end{document}